\documentclass{appolb}
\usepackage{graphicx,color}
\usepackage{amsmath}
\usepackage{float}
\usepackage{subcaption}
\usepackage{hyperref}
\usepackage{cite}

\begin{document}
\title{Screening masses towards chiral limit
\thanks{Presented at Criticality in QCD and the Hadron Resonance Gas, 29-31 July 2020,  Wroc\l aw, Poland.}%
}
\headtitle{Screening masses towards chiral limit}
\author{Simon Dentinger\thanks{Speaker}, Olaf Kaczmarek, Anirban Lahiri
	\address{Fakult\"at f\"ur Physik, Universit\"at Bielefeld, D-33615 Bielefeld, Germany}
}
\headauthor{S. Dentinger et al.}
\maketitle
\begin{abstract}
A possible effective restoration of the anomalous $U_A(1)$ symmetry would have a non-trivial effect on the global phase diagram of QCD. In this work we investigate the effective restoration of the $U_A(1)$ through the calculation of scalar and pseudo-scalar screening masses and corresponding susceptibilities, for physical and lower than physical pion masses. Calculations have been performed in (2+1)-flavor HISQ discretization scheme with a physical value of the strange quark mass. Preliminary calculations of the continuum extrapolated scalar and pseudo-scalar masses are presented, based on lattices with three different temporal extent. Non-trivial structure of the difference between scalar and pseudo-scalar susceptibilites are discussed for $N_\tau=8$ lattices.
\end{abstract}
  
\section{$U_A(1)$ symmetry and screening masses}
\label{sec:physical}


One important aspect for understanding the global QCD phase diagram is the fate of the anomalous $U_A(1)$ symmetry at the chiral phase transition, in the chiral limit $m_l \rightarrow 0$, for vanishing chemical potential $\mu_B = 0$. If $U_A(1)$ remains broken in the chiral limit then the chiral transition is expected to be of second-order belonging to the $O(4)$ universality class \cite{remark_chiral}. On the other hand if the $U_A(1)$ gets effectively restored at the chiral phase transition then the later can be of first-order \cite{remark_chiral} or second-order belonging to other universality classes \cite{axial_transition}. Therefore the study of effective restoration of the anomalous $U_A(1)$ towards the chiral limit is very important to decide the nature of the chiral phase transition. Existing calculations using staggered fermions \cite{screening_mass_paper,Cheng:2010fe,Ohno:2012br},
overlap and M\"obius domain wall fermions \cite{Buchoff:2013nra,Bazavov:2012qja,Tomiya:2016jwr}
as well as Wilson fermions \cite{Brandt:2016daq}  
are still inconclusive on the effective restoration of the anomalous $U_A(1)$ symmetry.
In this contribution we report on an attempt to study the breaking or restoration of the anomalous $U_A(1)$ through the degeneracy between pseudo-scalar (iso-vector) and scalar (iso-vector) meson channels which are related by $U_A(1)$ symmetry, towards the chiral limit. The explicit breaking of $U_A(1)$ through the quark mass term provides a background which is supposed to vanish in the chiral limit, i.e. $m_l \rightarrow 0$.

\section{Screening masses for lower than physical quark masses}
\label{sec:lower_physical}
We have calculated screening correlators in (2+1)-flavor QCD using the highly improved staggered quark (HISQ) action and tree level improved Symanzik gauge action. Screening mass calculations have been done following ref.\ \cite{screening_mass_paper}, where details about the staggered correlation functions and fit ansatz for different channels can be found. To approach the chiral limit, we have decreased the light quark masses keeping the strange quark mass fixed at its physical value. For details of the lattice setup see \cite{chiral_tc}.
Since a direct calculation at the chiral limit $m_l = 0$ is not possible on the lattice, to obtain  screening masses and the susceptibilities in the chiral limit, an extrapolation in mass apart from  thermodynamic and continuum limit extrapolations would be required. It is known that the staggered formulation at any finite lattice spacing preserves only a subgroup of the continuum chiral symmetry, thus it is necessary to take the continuum limit prior to the chiral extrapolation. To control the potential finite volume effects we need to take the thermodynamic limit first. In the following subsections we discuss these limits in detail. 

\subsection{Thermodynamic extrapolation}
The thermodynamic limit of screening masses can be taken using the following form \cite{finite-size}
\begin{equation}
	m_{N_s/N_\tau} = m_{N_s \rightarrow \infty / N_\tau} \bigg ( 1 + b_{N_\tau} \bigg( \frac{N_\tau}{N_s} \bigg)^c \bigg),
\end{equation}
where $m_{N_s/N_\tau}$ is the screening mass calculated on a $N_s^3\times N_{\tau}$ lattice. $m_{N_s \rightarrow \infty / N_\tau}$, being the mass in the thermodynamic limit, is a fit parameter along with $c$ and $b_{N_\tau}$. It has been argued that \cite{finite-size} $c=3$ for $T = 0$ and $c=1$ for $T \rightarrow \infty$. Therefore it is assumed here that $c \in [1,3]$ for any $T$. Assuming that $c$ only depends on temperature $T$ and the number of temporal lattice points $N_\tau$, while $m_{N_s \rightarrow \infty / N_\tau}$ and $b_{N_\tau}$ depend on $N_\tau$, channels ({\it e.g.}\ PS, S) and temperature $T$, a combined fit with a shared parameter $c$ between pseudo scalar and scalar particles at fixed $N_\tau$ and $T$ is possible.

\begin{figure}[thbp]
	\begin{subfigure}[t]{0.49\linewidth}
		\includegraphics[width=1\linewidth]{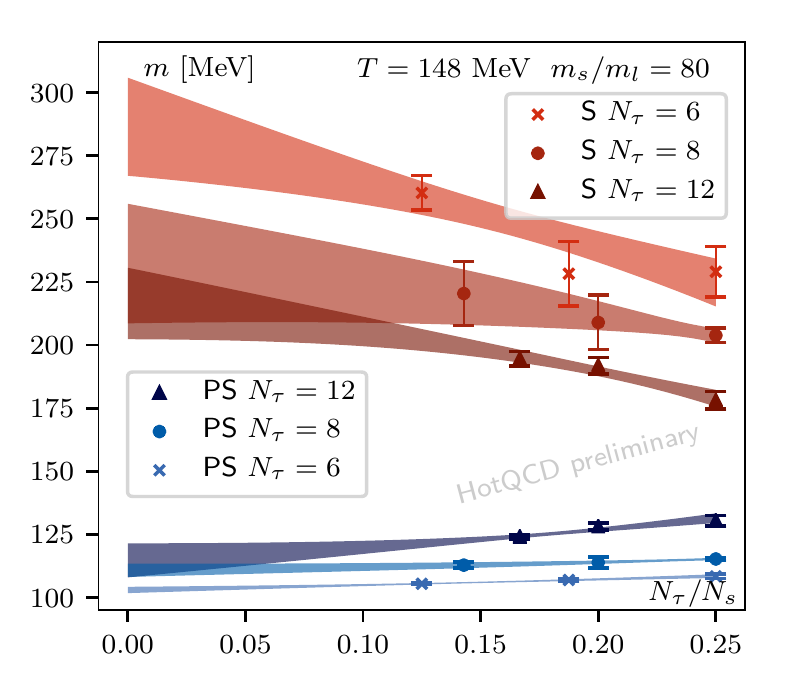}
	\end{subfigure}
	\begin{subfigure}[t]{0.49\linewidth}
		\includegraphics[width=1\linewidth]{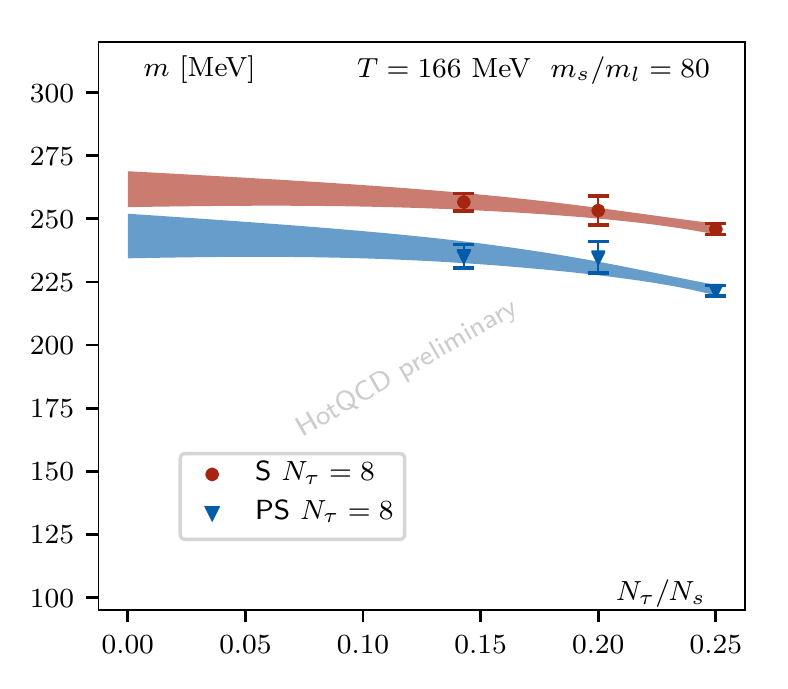}
	\end{subfigure}
	\caption{Infinite volume extrapolations of scalar (S, red) and pseudo scalar (PS, blue) channels at $T\approx 148$ MeV (left) and $T\approx 166$ MeV (right) for mass ratio $m_s/m_l=80$.}
	\label{fig:fitparamt148t147t148mr80ll}
\end{figure}

In Fig.\ \ref{fig:fitparamt148t147t148mr80ll} we show the screening masses in the scalar (S) and pseudo-scalar (PS) channels for three volumes at different $N_\tau$ together with the combined 
thermodynamic limit extrapolations at two temperatures. For the lower temperature volume effects are visible, but the screening masses for the largest volumes in the plot agree with the infinite volume extrapolated values within errors in most cases. For this temperature we have shown the volume extrapolation for 3 different $N_{\tau}$. For each $N_{\tau}$ we have performed the volume extrapolation jointly to S and PS masses. For the higher temperature case the two largest volumes  agree very well with the extrapolation results which in turn shows that volume effects are smaller for higher $T$ as also seen for other chiral observables \cite{chiral_tc}. From the fits the parameter $c$ comes out to be $1.33(66)$, $1.89(97)$ and $1.83(92)$ for $T\approx 148$ MeV at $N_\tau=6,8,12$ respectively and $2.59(76)$ for $T\approx 166$ MeV. Fig. \ref{fig:fitparamt148t147t148mr80ll} suggests that the use of the screening masses from the largest available volumes is well justified for temperatures close to the pseudo critical temperature and above.
This is further confirmed in the following section where we perform the continuum extrapolation at lower temperature of Fig.\ \ref{fig:fitparamt148t147t148mr80ll} based on the screening mass results on the largest volume in comparison to the screening masses obtained in the thermodynamic limit. 

\subsection{Continuum extrapolation}

\begin{figure}[thbp]
	\begin{subfigure}[t]{0.49\linewidth}
		\includegraphics[width=1\linewidth]{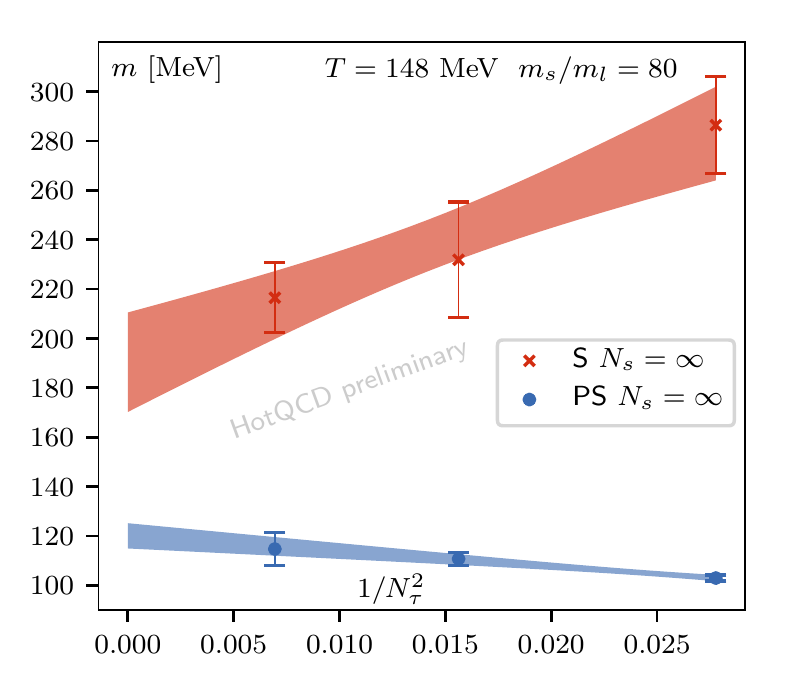}
	\end{subfigure}
	\begin{subfigure}[t]{0.49\linewidth}
		\includegraphics[width=1\linewidth]{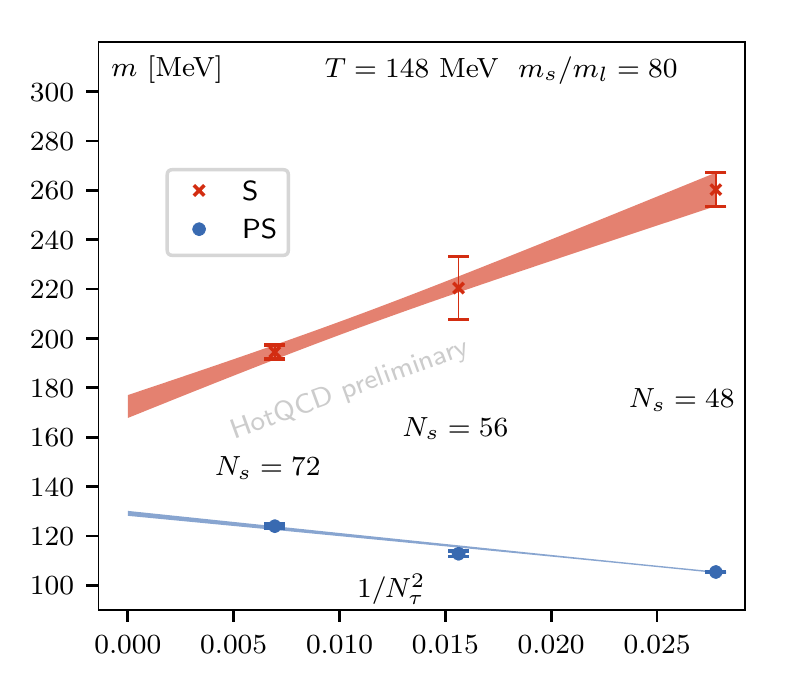}
	\end{subfigure}
	\caption{Continuum extrapolation of scalar (S, red, upper) and pseudo scalar (PS, blue, lower) channels for $T\approx 148$ MeV for mass ratio $m_s/m_l=80$. Left: Thermodynamic limit of the screening masses have been taken beforehand for each $N_{\tau}$, see left plot of Fig.\ \ref{fig:fitparamt148t147t148mr80ll}. Right: Screening masses from largest volume only.}
	\label{fig:fitparamcontt148t147t148mr80ll}
\end{figure}

\begin{figure}[thbp]
	\begin{subfigure}[t]{0.49\linewidth}
		\includegraphics[width=1\linewidth]{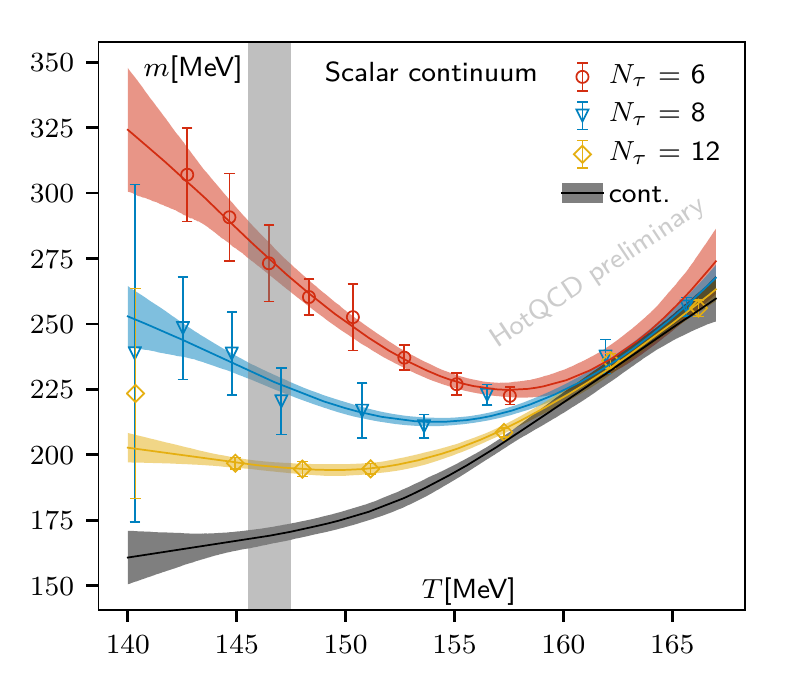}
	\end{subfigure}
	\begin{subfigure}[t]{0.49\linewidth}
		\includegraphics[width=1\linewidth]{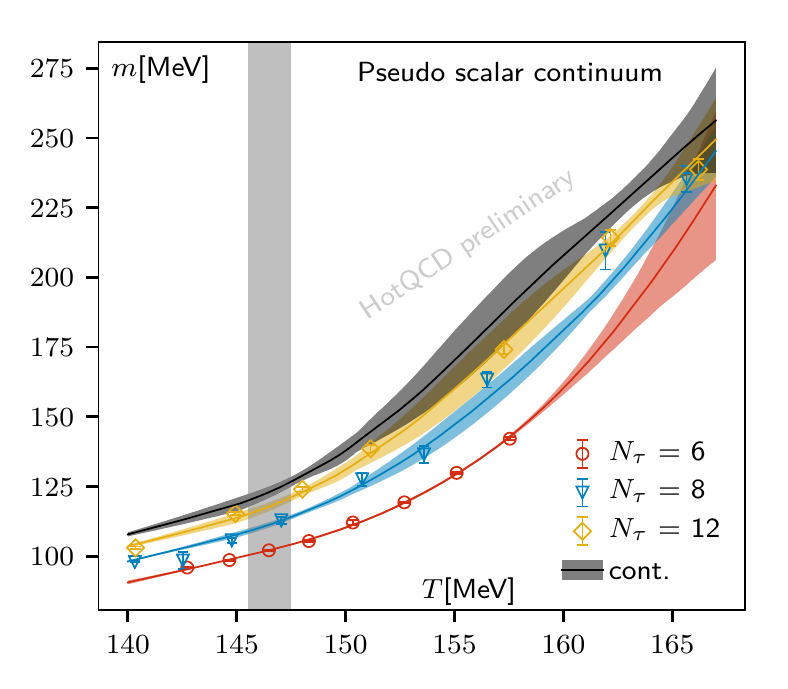}
	\end{subfigure}
	\caption{Continuum extrapolation for screening masses over the entire available temperature range for $m_s/m_l = 80$. Left: Scalar (S). Right: Pseudo-scalar (PS). 
	}
	\label{fig:extr_M1llmr80_test}
\end{figure}

In Fig.\ \ref{fig:fitparamcontt148t147t148mr80ll} we show the continuum limit extrapolation for $T\approx 148$ MeV with a linear ansatz in $1/N_\tau^2$. In the left panel we show the continuum limit extrapolation of the infinite volume extrapolated screening masses, obtained from Fig.\ \ref{fig:fitparamt148t147t148mr80ll} and in the right panel we show the continuum extrapolation of the screening masses from the largest available volumes. Our preliminary calculations give $m_{S} = 191(19)$ MeV and $m_{PS} = 120(5)$ MeV, in the continuum limit when we use the infinite volume extrapolated masses (left panel). Using the masses from the highest volumes (right panel) the continuum results are $m_{S} = 173(5)$ MeV and $m_{PS} = 129(1)$ MeV. Since the two continuum extrapolations give results which agree within 95\% confidence interval, for further analyses we use the screening masses from the largest available volumes.\\
Since the temperature values for different $N_\tau$ do not exactly match always, temperature interpolations are needed to have the continuum extrapolation for the entire temperature range available. Here we have used cubic splines with node positions determined by the density of the data points \cite{screening_mass_paper}.  These temperature interpolations for $N_\tau = 6,~8~\text{and}~12$ were used in the joint continuum extrapolation with spline coefficients being linear in $1/N_\tau^2$. Details about the continuum extrapolations can be found in ref.\ \cite{screening_mass_paper}.\\
In Fig. \ref{fig:extr_M1llmr80_test} we show preliminary results of this continuum limit extrapolation in the full available temperature range for mass ratio $m_s/m_l = 80$.  At low temperatures, $T = 25~\mathrm{MeV}$ and $T = 50~\mathrm{MeV}$, the fit is constrained on the first derivative to be equal to zero. For high temperature, well above the available temperature region of our study, a constraint at $T = 1.5~\mathrm{GeV}$ on the first derivative to be equal to $2 \pi$ has been used.
The gray band shows a continuum estimate of the pseudo-critical temperature for $m_s/m_l=80$ of 
$146.5 \pm 1.0~\mathrm{MeV}$ based on \cite{tpc_80}.

\begin{figure}[H]
	\centering
	\includegraphics[width=0.49\linewidth]{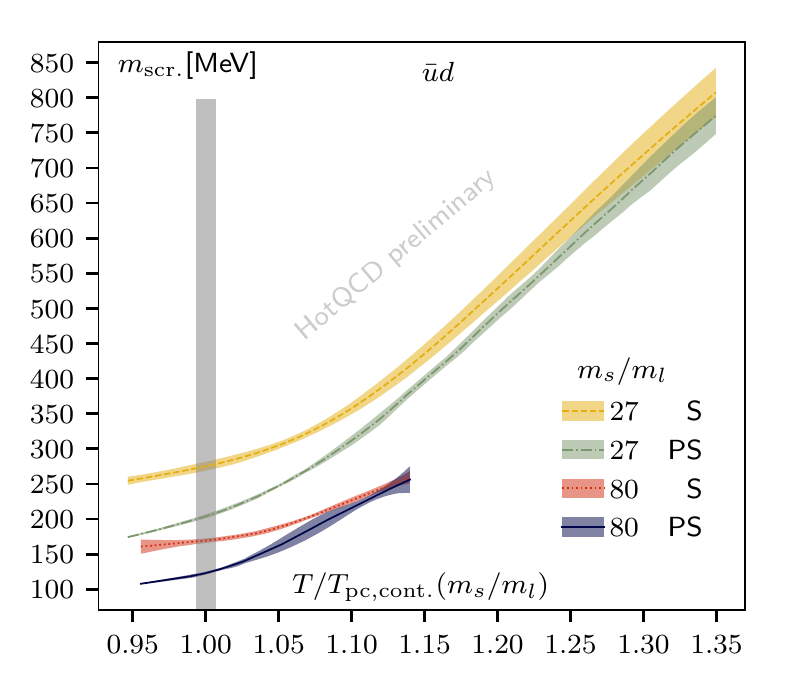}
	\caption{Comparison of the continuum extrapolated screening masses for $m_s/m_l=27$ and $m_s/m_l=80$ as function of temperature scaled by the respective pseudo-critical temperatures in the continuum limit.}
	\label{fig:screeningmasscontbandscompareH27H80}
\end{figure}
It is well known that for low temperatures ($T<T_{pc}$) the scalar channel suffers from unphysical decay modes in the staggered discretization. For a detailed discussion we refer the reader to ref.\ \cite{screening_mass_paper} and references therein. Above $T_{pc}$ this problem is expected to be mild or absent and one may use the degeneracy between the S and PS channel masses to explore the effective restoration of the anomalous $U_A(1)$ symmetry.
We show the continuum extrapolated results for the scalar (S) and pseudo-scalar (PS) channels for two quark mass ratios, $m_s/m_l=27$ (taken from \cite{screening_mass_paper}) and $m_s/m_l=80$,
in  Fig.\  \ref{fig:screeningmasscontbandscompareH27H80}. One can see that the scalar and pseudo-scalar meson masses become degenerate which indicates that $U_A(1)$ is effectively restored around the temperature region of $T \approx 160-175$ MeV corresponding to $T/T_{pc}\sim 1.1-1.2$ for $m_s/m_l=80$. Although a better controlled continuum extrapolation and results at higher temperatures would be helpful in this respect, at this point it is interesting to compare the situation to physical pion mass, $m_s/m_l=27$, in Fig.\ \ref{fig:screeningmasscontbandscompareH27H80}. One can see that the scalar and pseudo-scalar masses start becoming degenerate at about 200 MeV corresponding to $T/T_{pc}\sim 1.3$ for $m_s/m_l=27$. Primarily it might be tempting to conclude that with decreasing pion mass the degeneracy between S and PS channel shifts to lower $T/T_{pc}$ and results for lighter quark masses and a chiral extrapolation will be important to fully study this question in the future.

\section{Susceptibilities}
Keeping in mind the issue in the scalar channel mass related to the staggered discretization, it might be a better way to investigate $U_A(1)$ by using the integrated correlation functions (or susceptibilities) for scalar and pseudo-scalar mesons directly.
Susceptibilities for pseudo-scalar ($\pi$) and scalar ($a_0$) channel for staggered discretization are defined as
\begin{equation}
\chi_\pi=\sum_{n=0}^{N_s-1}{\cal G}_{M2}(n)
\quad\text{and}\quad
\chi_{a_0}= - \sum_{n=0}^{N_s-1}(-)^{n}{\cal G}_{M1}(n)
\end{equation}
respectively. Here $n$ denotes the distance between source and sink in lattice units and for the definition of the staggered correlators ${\cal G}_{M2}$ and ${\cal G}_{M1}$, we refer the reader to ref.\ \cite{screening_mass_paper}.
An advantage of susceptibilities is that they are independent of multiple state fits which are used for extracting screening masses. It has been shown earlier \cite{screening_mass_paper} that with the HISQ action for physical pion mass the degeneracy between the pseudo-scalar ($\chi_\pi$) and the scalar ($\chi_{a_0}$) susceptibilities happens around the degeneracy temperature for masses, which may not be too surprising because the restoration of a symmetry should be imprinted in the degeneracy of the corresponding correlators and therefore also in their large distance decay that determines the screening masses.
\begin{figure}[thbp]
	\centering
	\includegraphics[width=0.62\linewidth]{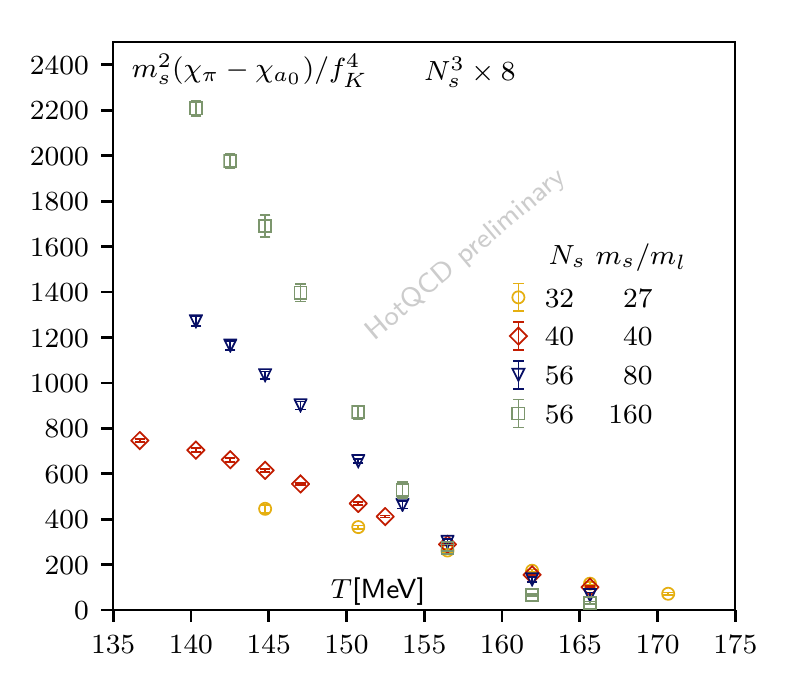}
	\caption{Temperature variation of $\chi_\pi$ and $\chi_{a_0}$ for various values of $m_l/m_s$ for $N_\tau=8$ lattices.}
	\label{fig:chipiminuschideltant8Tvar}
\end{figure}
In Fig. \ref{fig:chipiminuschideltant8Tvar} we show the difference between the pseudo-scalar ($\chi_\pi$) and scalar ($\chi_{a_0}$) as a function of temperature for various values of $m_l/m_s$ for lattices with $N_\tau=8$. The susceptibilities are renormalized by $m_s^2$ and normalized by the kaon decay constant $f_K^4$ to make it dimensionless. 
At low temperatures, in the chirally broken phase, the rapid increase of the difference between $\chi_\pi$ and $\chi_{a_0}$ can be expected owing to the ward identity $\chi_\pi=\langle\bar{\psi}\psi\rangle_l/m_l$ and to the fact that in the chiral limit below the transition temperature the chiral order parameter $\langle\bar{\psi}\psi\rangle_l$ is a constant due to the spontaneous breaking of the chiral symmetry. Thus the leading mass dependence is expected to be proportional to $1/m_l$ in $\chi_\pi$. On the contrary for $\chi_{a_0}$, the leading contribution is expected to come due to the well known Goldstone effect which vanishes in the continuum limit for 2 light flavors of the staggered quarks \cite{Smilga:1995qf}. Although at finite lattice spacing there will be a finite contribution proportional to $1/\sqrt{m_l}$ in $\chi_{a_0}$ \cite{Smilga:1995qf}. 

\begin{figure}[thbp]
	\centering
	\includegraphics[width=0.62\linewidth]{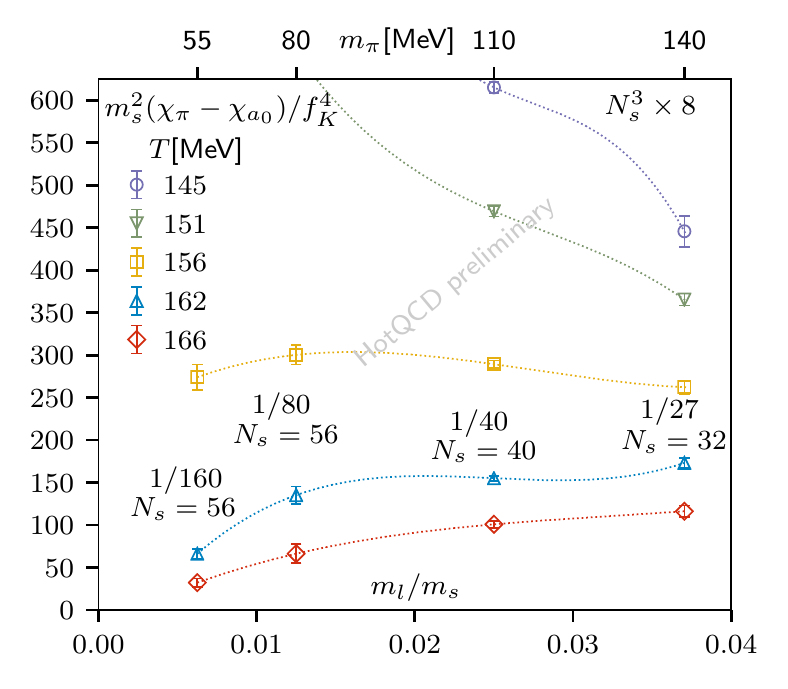}
	\caption{Variation of the difference between $\chi_\pi$ and $\chi_{a_0}$ with respect to quark mass for various values of temperatures for $N_\tau=8$ lattices. Dotted lines are cubic splines just for guiding the eye.}
	\label{fig:chipiminuschideltant8massvar}
\end{figure}

For temperatures around and above the chiral phase transition temperature, 
the mass dependence of the difference seems to be quite non-trivial. To show this more clearly, in Fig.\ \ref{fig:chipiminuschideltant8massvar} we show the variation of the difference between $\chi_\pi$ and $\chi_{a_0}$ for various temperatures. For the two highest temperatures the differences tend to approach zero towards the chiral limit, quite smoothly. As can be seen from Fig.\ \ref{fig:chipiminuschideltant8massvar}, for $T=156$ MeV the difference slightly increases and then seems to decrease towards chiral limit. This increase becomes more evident when the temperature further decreases and approaches $T_c$ and then this turning around happens for even lower masses. So the most non-trivial task is to find that after turning whether the difference vanishes or stays finite in the chiral limit. In this respect it is worth to mention that in a chirally symmetric background the difference between $\chi_\pi$ and $\chi_{a_0}$ is exactly equal to the disconnected part of the chiral susceptibility \cite{Kaczmarek:2020sif} and the behaviour of the later is found to be almost the same around the temperature regime shown in Fig.\ \ref{fig:chipiminuschideltant8massvar}.
Of course, it will be an interesting task to repeat the analyses of Fig.\ \ref{fig:chipiminuschideltant8Tvar} and Fig.\ \ref{fig:chipiminuschideltant8massvar} for other $N_\tau$ and then to take a continuum limit. Work is ongoing in that direction and will be reported elsewhere.

\section{Summary and Outlook}

In this contribution we have presented the calculations of screening masses in the pseudo-scalar (iso-vector) and scalar (iso-vector) channels within (2+1)-flavor HISQ discretization scheme with strange quark mass being fixed at its physical value and for the physical and lower than physical values of the (degenerate) light quark masses. Since for the lighter than physical pions, the finite volume effect is one of the main concern, we first discussed the finite volume effects for screening masses and also discussed the extrapolation to the thermodynamic limit. It is shown that for most of the temperatures under consideration the screening masses for the largest available volumes agree with the same in the thermodynamic limit within uncertainties. Finite volume effects for higher (compared to $T_{\text{pc}}$) temperatures are found to be even smaller. Furthermore we show that the PS and S masses in the continuum limit with infinite volume extrapolated masses or using masses from the highest available volumes for different $N_\tau$, match within 95\% confidence interval. Due to this small systematic uncertainty originating from the thermodynamic extrapolations we can safely proceed with continuum extrapolation of masses available on largest volumes. Then we compare the continuum extrapolated screening masses in PS and S channels for $m_s/m_l=27$ and preliminary results for $m_s/m_l=80$ to understand whether there exists any trend in degeneracy temperature with decreasing pion mass. Results for even lighter quark mass will help to understand the situation better in the future. As the staggered scalar channel suffers from unphysical decay modes, we also presented calculations of susceptibilites (for $N_\tau=8$) for PS and S channels and their difference as a measure of the $U_A(1)$ breaking. Preliminary calculations show that the degeneracy of the PS and S susceptibilites occur around the degeneracy temperature of the corresponding screening masses. Although a non-trivial behaviour of the difference between scalar and pseudo-scalar susceptibilities makes it harder to make definite comments about chiral limit. More exploration in this direction is needed and planned for the future.  


\section*{Acknowledgments}

This work was supported by the Deutsche Forschungsgemeinschaft (DFG, German Research Foundation) - Project  number 315477589-TRR 211 and the grant number 05P18PBCA1 of the German Bundesministerium
f\"ur Bildung und Forschung.


\end{document}